\documentclass[reprint,twocolumn,amsmath,amssymb,aps,pre,nofootinbib]{revtex4-1}
\usepackage[utf8]{inputenc}
\usepackage[T1]{fontenc}
\usepackage[english]{babel}
\usepackage{graphicx}
\usepackage{bm}
\usepackage[usenames,dvipsnames]{color}
\usepackage{hyperref}
\hypersetup{citecolor=ForestGreen, linkcolor=ForestGreen, urlcolor=CadetBlue, colorlinks=true}

\newcommand{\bq}{\mathbf{q}}

\newcommand{\id}{\mathrm{d}}
\newcommand{\br}{\mathbf{r}}

\newcommand{\bG}{\mathbf{G}}
\newcommand{\bg}{\mathbf{g}}

\newcommand{\bff}{\bm{f}}

\newcommand{\la}{\langle}
\newcommand{\ra}{\rangle}
\newcommand{\bnabla}{\bm{\nabla}}

\begin{document}
\title{Collective wake-mediated interactions in lattice fluids:\\
Effects of strong local force fluctuations stimulated by impurity disorder
}

\author{O.V.~Kliushnichenko} \email{kliushnychenko@iop.kiev.ua}
\affiliation{Institute of Physics, NAS of Ukraine, Prospect Nauky 46, 03028 Kiev, Ukraine}

\author{S.P.~Lukyanets} \email{lukyan@iop.kiev.ua}
\affiliation{Institute of Physics, NAS of Ukraine, Prospect Nauky 46, 03028 Kiev, Ukraine}

\begin{abstract}
The perturbation of a medium field (a particle wake) determines the Stokes drag on the particle itself. Besides, it affects the motion of other particles, inducing non-equilibrium correlations between them. We study the induced non-equilibrium correlations and forces, acting on constituent particles of the bunch or cluster of impurities exposed to a gas stream. Such induced correlations exhibit many striking features that are determined by the properties of particle bunch collective scattering in a medium.
The characteristics of collective scattering are determined by the structure of the bunch itself, that is associated with scattering on inhomogeneities, i.e., on fluctuations of the number of particles (scatterers) in a correlation volume. This is well-known, e.g., in optics and have been shown for gas scattering in our recent work \cite{kliushnychenko_effects_2018}. In particular, a random cluster of impurities experiences much stronger Stokes drag than a regular one.
Inhomogeneity of the cluster also determines the presence of giant local fluctuations of the scattered field inside a cluster, that was shown for gas stream scattering in \cite{kliushnychenko_effects_2018}. This, in turn, should lead to strong local fluctuations of induced gradient forces inside the cluster, which can determine its stability. Moreover, the description of a cluster in terms of effective parameters (penetration index, effective diffusion coefficient, etc.) brakes down due to the presence of such fluctuations.
\end{abstract}

\pacs{05.40.Jc, 47.70.Nd, 68.43.Jk}

\maketitle

\section{Motivation \& Model}

We focus on purely dissipative (diffusive) system and make use of the minimal classical two-component lattice gas model with hard-core repulsion: each lattice site can be occupied by only one particle. Despite the short range of inter-particle interaction it was shown to give rise to peculiar nonlinear effects essentially manifested at high gas concentrations: the dissipative pairing effects \cite{kliushnychenko_effects_2017,mejia-monasterio_bias-_2011}, the wake inversion and switching of wake-mediated interaction \cite{kliushnychenko_effects_2017}, formation of non-equilibrium structures \cite{vasilyev_2017} etc.

Kinetics of a two-component lattice gas is described by the standard continuity equation, $\dot n_i^\alpha=\sum_j\left(J^\alpha_{ji}-J^\alpha_{ij}\right)+\delta J_i^\alpha$, where $\alpha=1,2$ labels the particle species and $n_i^\alpha=0,1$ are the local occupation numbers of particles at the $i$th site. $J^\alpha_{ij}=\nu^\alpha_{ij}n_i^\alpha\left(1-\sum_\beta n_j^\beta\right)$ gives the average number of jumps from site $i$ to a neighboring site $j$ per time interval, $\nu_{ij}^\alpha$ is the mean frequency of these jumps. In what follows, fluctuations of the number of jumps (the term $\delta J_i^\alpha$) are neglected. To describe the scattering of particle stream by an impurity cloud we assume, see~\cite{kliushnychenko_effects_2017}, that one of the two components $u_i=0,1$ describes the given distribution of impurities and is static ($\nu_{ij}^1\equiv0$), while another one $n_i(t)$ is mobile. The presence of a weak driving field (force) $\bG$, $|\bg|=\ell|\bG|/(2kT)<1$ ($\ell$ is the lattice constant), leads to asymmetry of particle jumps for mobile component: $\nu_{ji}\approx\nu[1+\bg\cdot(\br_i-\br_j)/\ell]$. We use the mean-field approximation, $\partial_t\langle n_i\rangle=\sum_{j}(\langle J_{ji}\rangle-\langle J_{ij}\rangle)$, $\langle J_{ji}\rangle=\nu_{ji}\langle n_j\rangle(1-\langle n_j\rangle-u_i)$, where $\langle n_i\rangle=\langle n(\br_i)\rangle\in[0,1]$ describes the mean occupation numbers at sites $\br_i$ or the density distribution of flowing gas particles, $n_0\equiv n(|\br|\rightarrow\infty)$ being the equilibrium gas concentration (bath fraction). In what follows, we consider the two-dimensional (2D) case.

The macroscopic kinetics of the mobile component $n$ is given by the equation
\begin{equation*}\label{eq:lw}
  \partial_\tau n = \nabla^2n-\bnabla(u\bnabla n - n\bnabla u)-(\bg\cdot\bnabla)[n(1-u-n)],
\end{equation*}
where $n=n(\br,\tau)$ and $u=u(\br)$ are the average occupation numbers of the two components at the point $\br$ ($0\leq n\leq1$ and $0\leq u\leq1$) and $\bg=\ell\bG/(2kT)$.

We consider the properties of non-equilibrium formations resulting from scattering of gas stream by a cloud of impurities and examine the role of collective effects. We examine the effects of inner structure of impurity clusters, total drag (friction) force, accompanied by the nonlinear blockade effect in a gas.
We show that the nonlinear blockade effect considerably affects collective scattering.

\section{Collective Scattering Effects}

\begin{figure*}
\includegraphics[width=.8\textwidth]{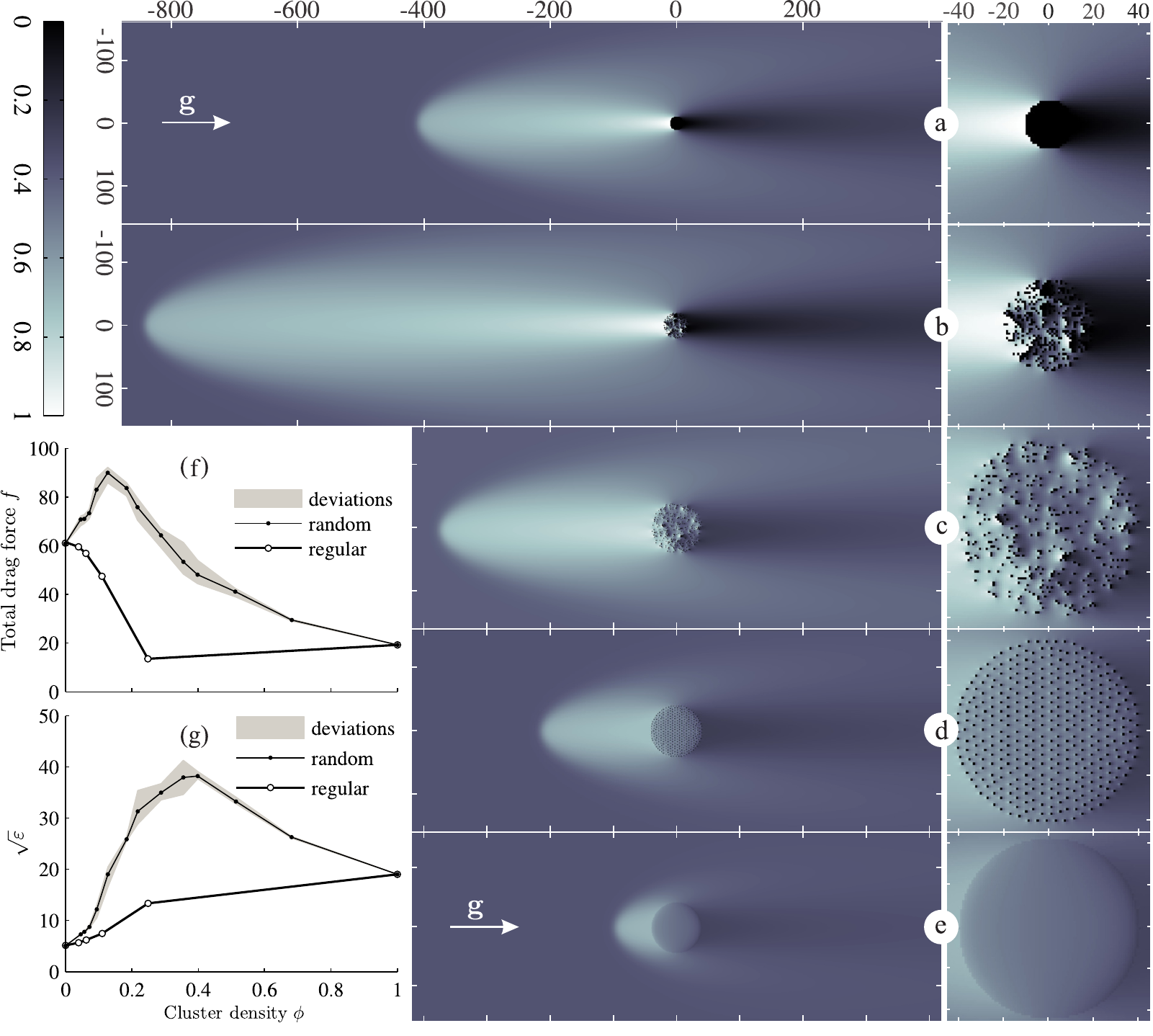}
\caption{\label{fig:cluster}
\textit{Collective enhancement of scattering.} Steady-state distributions of mean concentration $\la n(x_i,y_i)\ra$ [panels (a) to (e)] illustrate enhanced scattering (blockade region growth) for heterogeneously fractured obstacle. Coordinates are in units of $\ell$ (lattice constant). A close view of impurity cluster inner structure for each case is shown at the top: (a) solid obstacle, (b) and (c) random clusters, (d) regular cluster, (e) uniform cluster. $R=10.8\ell$ for (a), $R=20\ell$ for (b), and $R=40\ell$ for (c)--(e). Number of constituent single-site impurities is $N=362$, $n_0=0.37$, $|\bg|=0.5$ (stream is directed along the $x$-axis) for all calculated distributions. Plots (f) and (g) show dependencies of total drag force $f\equiv|\bff|$ (units of $kT/\ell$) and $\sqrt{\varepsilon}$ on cluster density $\phi$; $R\in(\infty\,\,10.8\ell]$, $N=362$, $n_0=0.2$.}

\end{figure*}

Two basic results are readily seen from Figs~\ref{fig:cluster}(a)--(e):
i) Fragmentation of a solid obstacle into a cluster of separate impurities considerably enhances the gas stream scattering.
ii) Enhancement of scattering is provoked by inhomogeneity of impurity distribution within a cluster; the scattering is less efficient for regularly ordered cluster, Figs~\ref{fig:cluster}(c)--(e) and (f).
This effect is analogous to that of light scattering on inhomogeneities in distribution of atoms (dipole moments) that is determined by the fluctuation of their number density in a definite volume or by the two-point correlation function.

The magnitude of scattered field $\delta n(\br)$ can be characterized by a quantity like total density dispersion $\varepsilon\equiv\overline{\delta n^2}\propto\int\delta n^2(\br)\,\id \br$. Figure~\ref{fig:cluster}(i) shows that dependence $\varepsilon(N)$ for impurity cluster can become power-law and, in particular, for random cluster is $\propto N^2$ that signifies the intrinsically collective scattering.

As Fig.~\ref{fig:cluster}(f) suggests, the dependence of total drag force, acting on impurity cluster, on its  density $\phi$ is qualitatively different for random and regular ones.

\section{Gradient (Wake-Mediated) Interaction within a Cluster}
In the linear approximation near equilibrium density $n_0$, so that $n=n_0+\delta n$, the Green function
\begin{equation*}\label{eq:Append:GreenFunc}
  G_{2(3)} = Q_{2(3)}(|\br-\br'|)e^{\bq\cdot(\br-\br')}, \quad \bq=(1/2-n_0)\bg,
\end{equation*}
is of the form of screened anisotropic Coulomb potential with $Q_3(r)=e^{-qr}/(4\pi r)$ in 3D and $Q_2(r) = K_0(qr)/(2\pi)$ (asymptotically $\sim qr^{-1/2}$ at $qr\gg1$) in 2D, see \cite{kliushnychenko_blockade_2014,kliushnychenko_effects_2017}.
This leads to anisotropic screening length $R_{\textrm{scr}}$ which behaves as $R_{\textrm{scr}}\sim [q(1-\cos\theta)]^{-1}$, where $q\equiv|\bq|$, $\theta$ is the angle between $\bg$ and $\br$.

The effective dissipative (wake-mediated) interaction between small and distant impurities, associated with this type potentials, belongs to the induced dipole-dipole (generally, multipole) interaction in the non-equilibrium steady state, and features the non-Newtonian (non-reciprocal) character \cite{kliushnychenko_effects_2017}.
Wake-mediated interaction is more pronounced when the collective wake (common density perturbation ``coat'' around impurities) is formed. These forces depend 
on gas concentration, magnitude of external driving flow and on mutual alignment of impurities \cite{kliushnychenko_effects_2017}.

\section{Giant Density Deviations inside Random Cluster of Impurities}
Disordered distribution of impurities can provoke strong local fluctuations of scattered field $\delta n(\br)$ inside a cluster, compared to average density: $\delta n^2(\br_i)>n_0^2$.
\vskip.5cm
\begin{figure}
\includegraphics[width=\columnwidth]{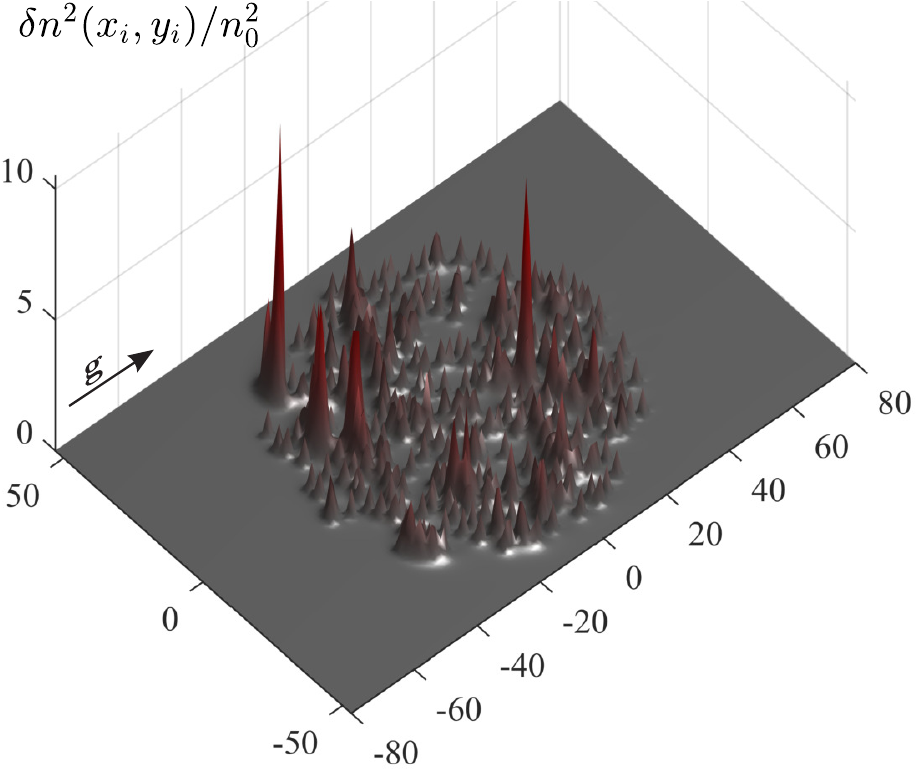}
\caption{\label{fig:density_fluct}
Strong local fluctuations of scattered field $\delta n(x_i,y_i)$ inside a random impurity cluster. $\phi=0.0569$ ($N=362$, $R=45\ell$), $n_0=0.2$, $|\bg|=0.5$.}
\end{figure}

\subsection{Number Fluctuations in Elementary Volume}
The enhancement of scattering can be provoked by inhomogeneity of scatterers distribution within an ensemble. On the one hand, this effect is determined by the fluctuation of number density of scatterers in a definite volume or by the two-point correlation function in analogy with light scattering on inhomogeneities in the distribution of atoms \cite{sobelman2002}.

On the other hand, the probability of meeting a second impurity near the given one is higher than the mean
probability of meeting it anywhere else. Thus, there appears another structural unit, a \textit{cluster} or a \textit{bunch} (say, ``cluster of galaxies'' \cite{zeldovich1965survey}, in astrophysical context).
Following Zeldovich \cite{zeldovich1965survey}, a different procedure can be exploited: The space is divided into separate volumes $V$ and the number of impurities in each volume is calculated. Then, as the quantitative characteristics of distribution, the root-mean-square deviation of the number of impurities in each volume is used:
$\overline{\delta N} = \sqrt{\frac{1}{k}\sum_{1}^{k}\left(N_k-\overline{N}\right)^2}$.
In the same manner, other associated quantities can be defined and calculated.

\section{Strong Local Variations of Forces}
\begin{figure}
\includegraphics[width=.85\columnwidth]{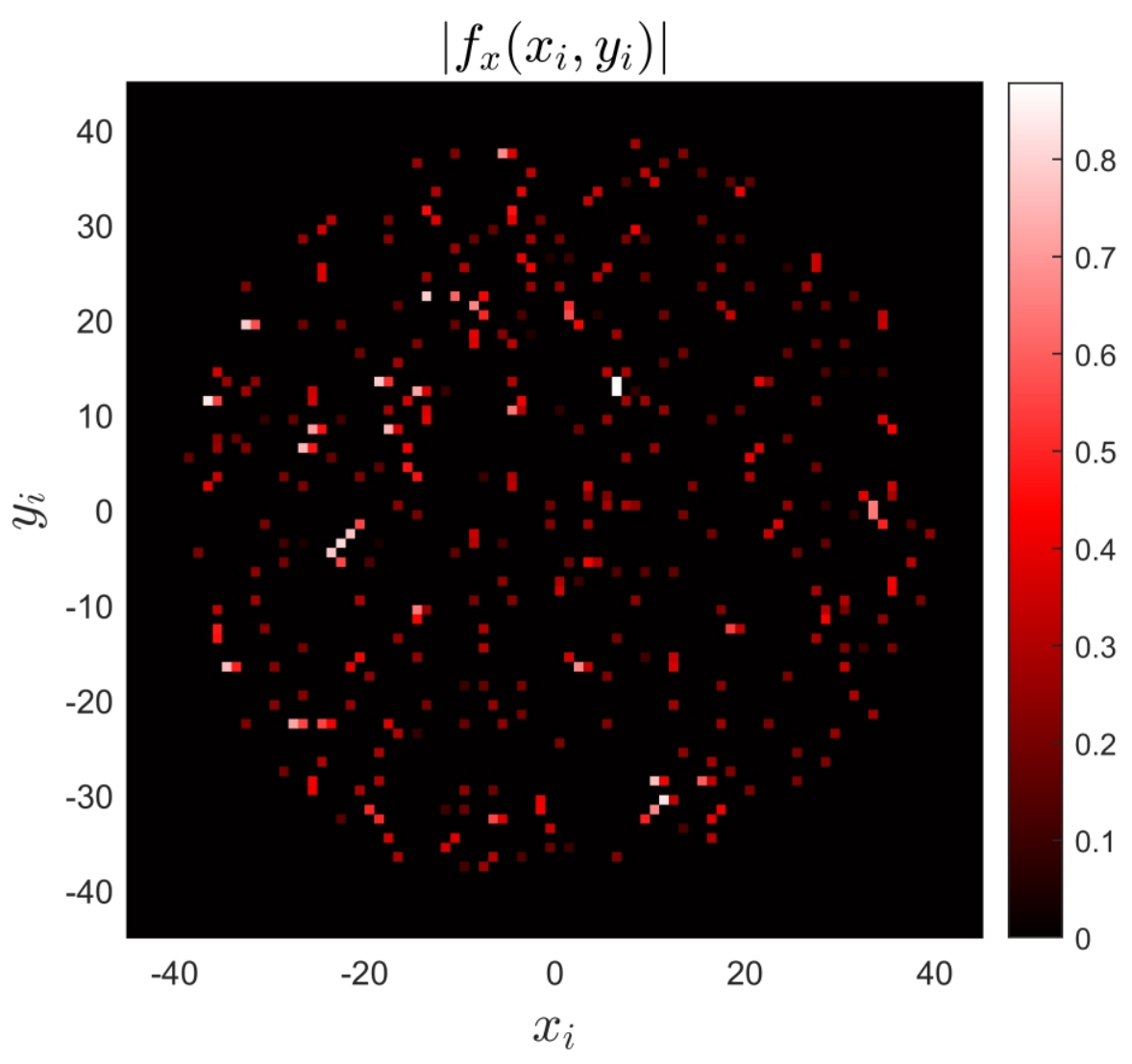}
\caption{\label{fig:reals}Spatial distribution of force magnitude $|f_x(x_i,y_i)|$ calculated numerically for disordered impurity cluster shown on Fig.~\ref{fig:cluster}(c). The ``lattice version'' of the expression for the force has been used, following the approach from \cite{mejia-monasterio_bias-_2011}.}
\end{figure}

Expectedly, strong local variations of scattered field (density fluctuations) illustrated by Fig.~\ref{fig:density_fluct}, lead to considerable fluctuations of forces on impurities. As can be seen from Fig.~\ref{fig:reals}, a little over a dozen of impurities (see white spots) experiences the highest values of drag force $\bff_{drag}\equiv f_x$.
\begin{figure}
\includegraphics[width=\columnwidth]{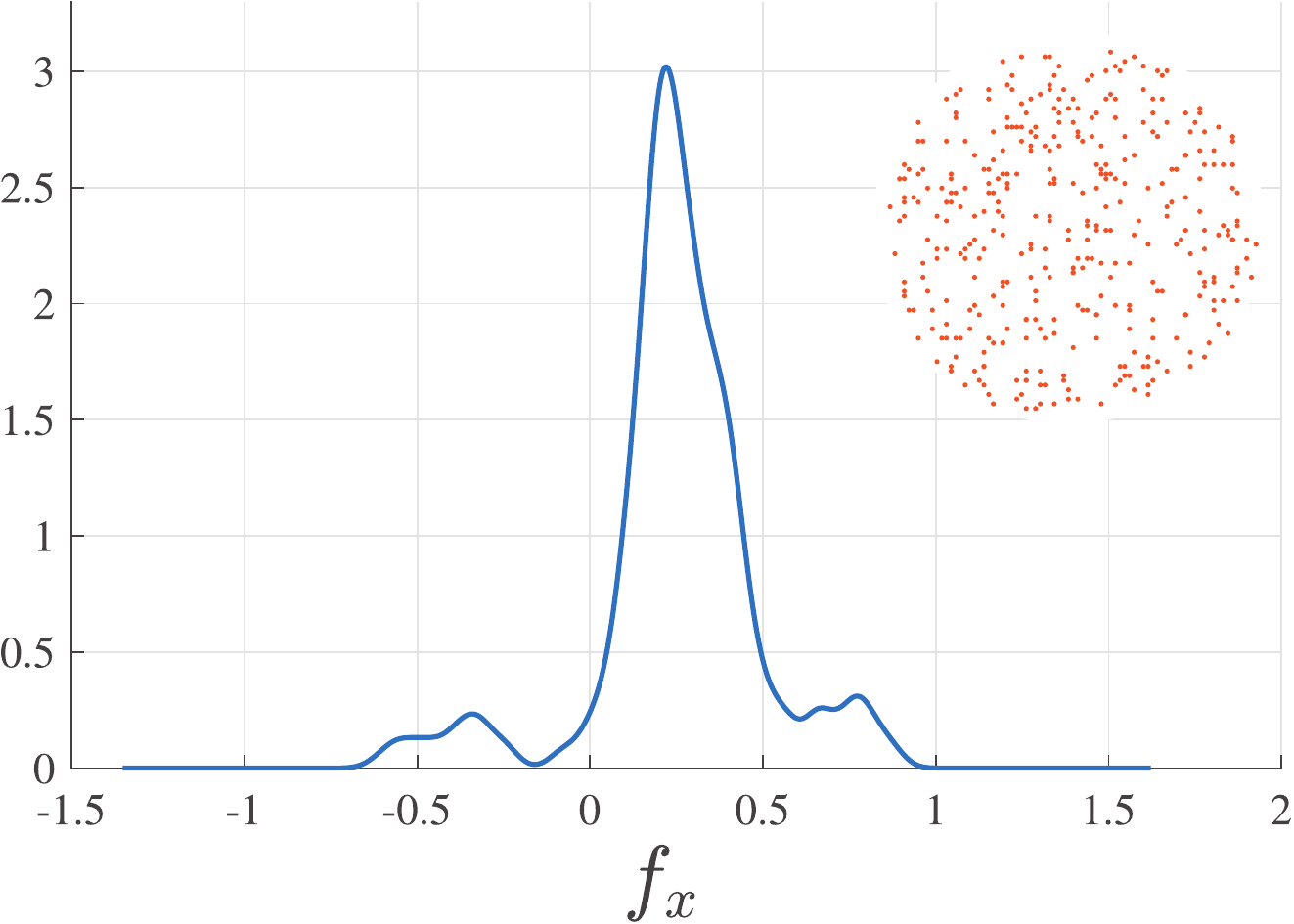}
\caption{\label{fig:tails}Probability density fitting for the force component $f_x$ distribution. Kernel density estimator developed in \cite{kernel} has been exploited. The inset shows the spatial distribution of impurities within the considered cluster.}
\end{figure}

\subsection{Role of Heavy Tails}
The heavy tails in the distribution of force values, Fig.~\ref{fig:tails} reflects the presence of local and rare but strong fluctuations of forces, acting on certain impurities in the cluster.

\section{Conclusions}
Certain features of induced non-equilibrium correlations and forces, acting on constituent particles of the bunch or cluster of impurities in a gas are shown. Those are determined by the properties of particle bunch collective scattering in a medium. The characteristics of collective scattering are determined by the structure of the bunch itself, that is associated with scattering on inhomogeneities, i.e., on fluctuations of the number of particles (scatterers) in a correlation volume. Inhomogeneity of the cluster also determines the presence of giant local fluctuations of the scattered field inside a cluster. This, in turn, lead to strong local fluctuations of forces on impurities inside the cluster.
The fat-tailed distribution of forces can be attributed to the presence of infrequent extreme deviations, as opposed to frequent modestly sized deviations. This can be essential for impurity cluster stability since certain impurities within a cluster experience the action of strong force which value considerably exceeds the mean force value in the cluster.

\acknowledgments{O.K. was supported by a grant for research groups of young scientists from the National Academy of Science of Ukraine (Project No. 0120U100155).}


\begin{thebibliography}{99}

\bibitem{kliushnychenko_effects_2018}
O.~V.~Kliushnychenko, and S.~P.~Lukyanets, \textit{Effects of collectively induced scattering of gas stream by impurity ensembles: Shock-wave enhancement and disorder-stimulated nonlinear screening}, Phys. Rev. E \textbf{98}, 020101(R) (2018).

\bibitem{kliushnychenko_effects_2017}
S.P. Lukyanets, O.V. Kliushnychenko, Phys. Rev. E \textbf{95} 012150 (2017)

\bibitem{mejia-monasterio_bias-_2011}
C.~Mej{\'i}a-Monasterio, G.~Oshanin,
Soft Matter \textbf{7}, 993 (2011)

\bibitem{vasilyev_2017}
O.A.\,Vasilyev \textit{et al.},
Soft Matter \textbf{13}, 7617 (2017)

\bibitem{kliushnychenko_blockade_2014}
O.~V.~Kliushnychenko and S.~P.~Lukyanets,
J. Exp. Theor. Phys. \textbf{118}, 976 (2014).

\bibitem{kernel}
Z.I. Botev, J.F. Grotowski, and D.P. Kroese, \textit{Kernel density estimation via diffusion}, The annals of Statistics \textbf{38}, 2916 (2010).

\bibitem{sobelman2002}
I.I. Sobel’man, Phys. Usp. \textbf{45}, 75 (2002).

\bibitem{zeldovich1965survey}
Ya. B. Zeldovich, \textit{Survey of modern cosmology} (in \textit{Advances in astronomy and astrophysics}, Vol.~3, P.~241--379), (Elsevier, 1965)

\end{thebibliography}
\end{document}